# High flux lithium antineutrino source with variable hard spectrum

**V I Lyashuk[1,2]**

[1]Institute for Nuclear Research, Russian Academy of Sciences, 117312 Moscow, Russia
[2]National Research Center Kurchatov Institute, 123182, Moscow, Russia

The work is devoted to development of hard $\tilde{\nu}_e$-source with variable and regulated (controlled) spectrum. The high flux antineutrino source with hard $\tilde{\nu}_e$-spectrum based on neutron activation of $^7$Li and subsequent fast $\beta^-$-decay ($T_{1/2}$ = 0.84 s) of the $^8$Li isotope with emission of $\tilde{\nu}_e$ with energy up to 13 MeV - is discussed. Creation of the intensive isotope neutrino source of hard spectrum will allow increase the detection statistics of neutrino interaction and is especially urgent for oscillation experiments. The scheme of the proposed neutrino source is based on the continuous transport of the created $^8$Li to the neutrino detector, which moved away from the place of neutron activation. Analytical expressions for lithium antineutrino flux is obtained. The discussed source will ensure to increase the cross section for $(\tilde{\nu}_e,d)$-reactions from several times to tens compare to the reactor $\tilde{\nu}_e$-spectrum. An another unique feature of the installation is the possibility to vary smoothly the hardness of the neutrino spectrum.

## 1. INTRODUCTION. LITHIUM ANTINEUTRINO SPECTRUM. GENERALIZED HARDNESS OF THE TOTAL NEUTRINO SPECTRUM

The results in registration, accumulation and understanding of neutrino data in many cases are determined by characteristics of the neutrino source, such as neutrino flux and spectrum. First of all the main problems of neutrino detection are associated with extremely small cross sections of these reactions. So, cross sections of the $(\tilde{\nu}_e,p)$, $(\tilde{\nu}_e,e^-)$ and $(\tilde{\nu}_e,d)$-interaction are in the interval $10^{-46}$-$10^{-43}$ cm$^2$ for "reactor" energy. The smallness of cross sections extremely complicates separation of neutrino events from background. What is why a high neutrino flux is determining requirement for obtaining of reliable results. The most intensive neutrino flux is ensured by nuclear reactors - the most widely and traditionally used $\tilde{\nu}_e$-sources.

In spite of the apparent superiority on neutrino flux the nuclear reactors has a disadvantage: too-small hardness of $\tilde{\nu}_e$-spectrum. This character is extremely negative as the probability of registration strongly depends on neutrino energy. For the considered here reactor antineutrino energy the neutrino cross section is proportional to it's energy squared - $\sigma_\nu \sim E_\nu^2$. But antineutrinos $\tilde{\nu}_e$ emitted at $\beta^-$-decay of fission fragments in a nuclear reactor have rapidly decreasing spectrum and energy $E_{\bar{\nu}} \leq$ 10 MeV. The neutrino spectrum from $^{235}$U (as the main fuel component) is presented in the figure 1 in comparison with $^8$Li neutrino spectrum [1,2].

The disadvantage of rapidly dropping spectrum can be filled having realized the idea to use a high-purified isotope $^7$Li [3] for construction of lithium blanket (also called as converter) around the active zone of a reactor [4]. A short-lived isotope $^8$Li ($T_{1/2}$ = 0.84 s) is created under reactor neutron flux in the reaction $^7$Li(n,$\gamma$)$^8$Li and at $\beta^-$-decay emits hard antineutrinos of a well determined spectrum with the maximum energy $E_{\bar{\nu}}^{max}$ =13.0 MeV and mean value $\bar{E}_{\tilde{\nu}}$ = 6.5 MeV. So, this blanket will act as a converter of reactor neutrons to antineutrinos. In fact the such construction of the blanket around the active zone (as a neutron source) is the most simple scheme of lithium antineutrino source. We can call this type of the $\tilde{\nu}_e$ source as steady spectrum source. As a result the total $\tilde{\nu}_e$-spectrum from the active zone of a reactor and from decays of $^8$Li isotope becomes considerably more harder in comparison with the purely reactor neutrino spectrum. Note that reactor antineutrino spectrum is specified also with another problem as instability in time due to dependence of partial spectra from nuclear fuel composition of $^{235}$U, $^{239}$Pu, $^{238}$U and $^{241}$Pu which vary in time in operation period. The distribution of the total reactor $\tilde{\nu}_e$ –spectrum is known with significant errors which strongly rise at the energy above ~6 MeV [5,6].

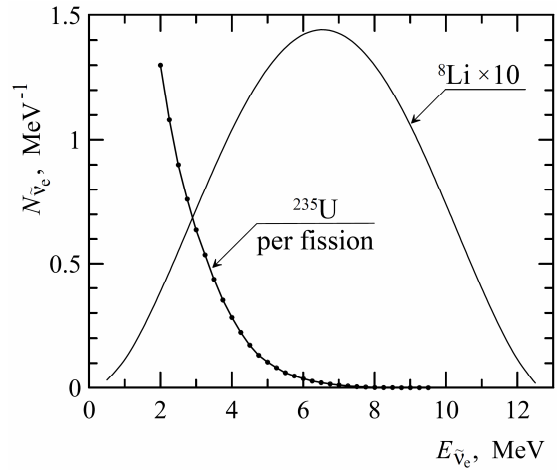

Figure 1. Spectrum of antineutrinos from $\beta^-$ -decay of $^8$Li [2] and fission fragments of $^{235}$U [1].

Let us define the productivity factor of the blanket $k$ (or possible to call $k$ as coefficient of blanket efficiency) as number of $^8$Li nuclei produced in the lithium per one fission in the active zone. It is clear that the hardness of the total spectrum will more larger as productivity factor will be more higher. An illustration of the resulting total spectrum in case of rising productivity factor is given in the figure 2. The range of questions connected with constructing and possible applications of intensive $\tilde{\nu}_e$ -sources (operating as steady one) of hard spectrum were considered in the works [4,6-10].

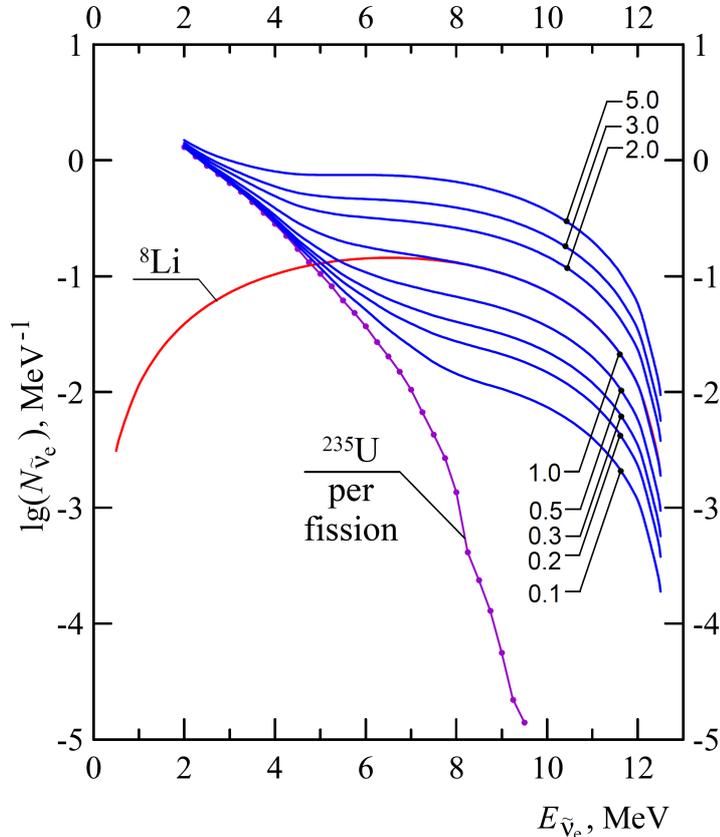

Figure 2. Neutrino spectrum from: $^{235}$U [1], $^8$Li [2] and total spectra from combination of active zone and lithiun blanket for different factors $k$ indicated for curves.

For our purpose (creation of the neutrino source of significantly larger hardness than possible to obtain by above mentioned simple scheme) let us introduce the definition of the generalized hardness for total neutrino spectrum. Let $F_{Li}(\vec{r})$ and $F_{AZ}(\vec{r})$ - densities of lithium antineutrinos flux from the blanket and antineutrino flux from the active zone, $\bar{n}_\nu = 6.14$ - number of reactor antineutrinos emitted per one fission in the active zone. We admit that the hardness of the summary $\tilde{\nu}_e$ - spectrum at the point $\vec{r}$ equals one unit of hardness if the ratio of densities $F_{Li}(\vec{r})/F_{AZ}(\vec{r})$ equals to $1/\bar{n}_\nu$. Then the total spectrum generalized hardness is:

$$H(\vec{r}) = \bar{n}_\nu \frac{F_{Li}(\vec{r})}{F_{AZ}(\vec{r})} \quad . \tag{1}$$

This definition is convenient as in so doing the averaged (over the blanket volume) value for the total $\tilde{\nu}_e$-spectrum generalized hardness of steady spectrum sources (these models are considered in [4,6,7]) is estimated by the value of its productivity factor $k$ of the blanket. Taking into account this definition of the hardness the values of productivity factors $k$ on the figure 2 coincide with values of the generalized hardness $H$ for total spectra.

## 2. LITHIUM ANTINEUTRINO SOURCE WITH VARIABLE AND REGULATED SPECTRUM

It is possible to supply powerful neutrino fluxes of considerably greater hardness by a facility with a transport mode of operation: liquid lithium is pumped over in a closed cycle through a blanket and further toward a remote neutrino detector (figure 3). For increasing of a part of hard lithium antineutrinos a being pumped reservoir is constructed near the $\tilde{\nu}_e$-detector. Such a facility will ensure not only more hard spectrum in the location of a detector but also an opportunity to investigate $\tilde{\nu}_e$-interaction at different spectrum hardness varying a rate of lithium pumping over in the proposed scheme with the closed loop [8,11].

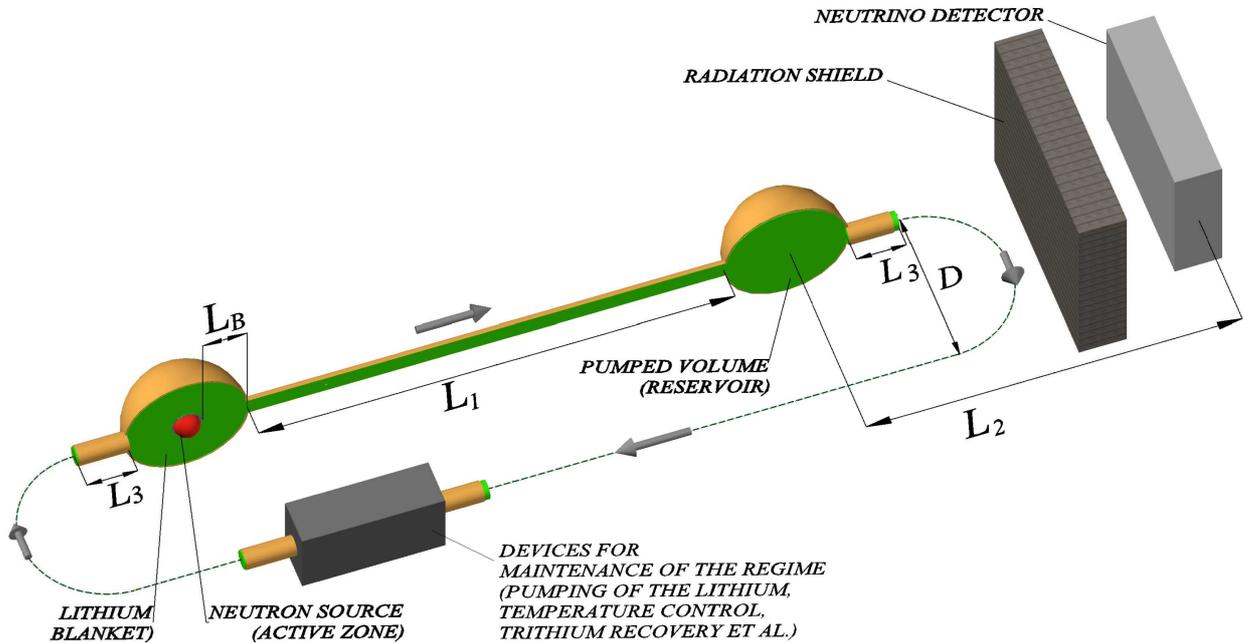

Figure 3. Scheme of the neutrino source with variable (regulated and controlled) spectrum. Lithium in the blanket (activated by neutrons from the source - reactor active zone) is pumped continuously through the delivery channel to the remote volume (reservoir, which is set close to the neutrino detector) and further back to the blanket. The rate of pumping can be smoothly varied by the installation for maintenance of the regime.

### 2.1. Problem of the large mass of purified lithium-7 isotope

The nature lithium consists of two isotope $^6Li$ and $^7Li$ with concentration 7.5% and 92.5% correspondingly. The beneficial $^7Li(n,\gamma)^8Li$ cross section is very small compare to large parasitic absorption on $^6$Li. So, at thermal point (neutron energy $E_n \sim 0.025$ eV) the cross section $\sigma_\alpha^{thermal}(^6\text{Li}) \approx$ 937 b, but this one for $^7$Li(n,γ)-activation is lower in four orders - $\sigma_{n\gamma}^{thermal} \approx 45$ mb. It is clear we need to ensure high grade of $^7$Li purification. Results of modeling show that requested $^7$Li purity must be about 0.9999 (or 0.9998 as minimum level). In order to fill the volume of the proposed scheme (figure 3) we will need more than 20 m$^3$ of lithium: the numerical data presented in the article below (in the part 3) correspond to 22 m$^3$, i.e., 12.2 t of $^7$Li isotope ($\rho = 0.553$ g/cm$^3$ [12]).

The main input to blanket productivity factor $k$ gives the thermal neutrons. So the logical solution is the next: to use material with high slowing-down power, but with very small absorption in order to ensure significant capture on $^7$Li. We propose to use such perspective substance as a heavy water solution of lithium hydroxides - $^7$LiOD, $^7$LiOD·D$_2$O [7]. Really this approach will help to solve two problem: 1) to pump a heavy water solution in the scheme with variable spectrum is more simple and safe compare to transport of metallic lithium; 2) the requested mass of high purified $^7$Li will strongly decrease (the price of installation will be heavy smaller). For our example: in case of 22 m$^3$ the mass of $^7$Li in heavy water solution (concentration of LiOD - 9.5 %) will be 0.87 t instead of 12.2 t; for smaller LiOD concentration (5.7%) the mass of $^7$Li will drop in 23.5 times (i. e., 0.52 t) at unimportant decrease of the blanket productivity factor $k$.

*2.2. Fluxes of lithium antineutrinos*

The creation of the $^8$Li at (n,γ)-activation in the blanket is described by equations:

$$\begin{cases} \dfrac{dN_7(t)}{dt} = -\lambda_{n\gamma} N_7(t) \\ \dfrac{dN_8(t)}{dt} = -\lambda_{n\gamma} N_7(t) - \lambda_\beta N_8(t) \end{cases} \quad (2)$$

where $N_7(t)$ and $N_8(t)$ - number of nucleus $^7$Li and $^8$Li at the time $t$, $\lambda_{n\gamma}$, $\lambda_\beta$ - rate of (n,γ)-reaction and $\beta^-$-decay.

We assume that at the instant $t = 0$ number of $^7$Li nucleuses (starting isotope) is $N_7^0$ and number of $^8$Li is $N_8^0 = 0$. The solution for system (2) is the next:

$$N_7(t) = N_7^0 \cdot \exp(-\lambda_{n\gamma} t) \quad (3)$$

$$N_8(t) = \lambda_{n\gamma} N_7^0 \frac{\exp(-\lambda_{n\gamma} t) - \exp(-\lambda_\beta t)}{\lambda_\beta - \lambda_{n\gamma}} . \quad (4)$$

The rapid decay of $^8$Li causes necessity of a maximally rapid delivery of lithium substance to a being pumped reservoir to ensure a greater hardness of the summary $\tilde{\nu}_e$ - spectrum in a location of the detector. Taking into account that $\lambda_{n\gamma} \ll \lambda_\beta$ and the time of (n,γ)-activation is a time of pumping over of a blanket volume and this time equals to several seconds (the time of pumping over is discussed in the part 3), the expression (4) is simplified:

$$N_8(t) = \frac{\lambda_{n\gamma} N_7^0}{\lambda_\beta} \left[1 - \exp(-\lambda_\beta t)\right]. \quad (5)$$

The value $\lambda_{n\gamma} N_7^0$ is the number of nuclei $^8$Li created in a time unit. On definition this value is the blanket productivity factor $k$ in view of the accepted normalization per one fission in the active zone. The knowledge of the $\lambda_{n\gamma} N_7^0$-value permits to calculate lithium $\tilde{\nu}_e$-fluxes from any parts of this cyclic system and the summary hardness of $\tilde{\nu}_e$ - spectrum in a location of the neutrino detector [13].

*2.2.1. Antineutrino flux from the lithium blanket*

Let us define the parameters of the source with variable spectrum (see figure 3): $V_B$ - blanket volume, $V_0$ - volume of the whole system, $w$ - rate of flow (i.e., volume being pumped over in a time unit), then $t_p = V_B / w$ - time of pumping over of the blanket volume. We will assume that transport time of all nuclei through the blanket is equal to $t_p$.

We consider the steady state mode of the source operation when fluxes escaping from different parts of the scheme do not vary in time. Then during the time $t_S = V_S / w$ (for pumping of some volume $V_S$ through the blanket) the part of $^8$Li nuclei created within this time interval $[0, t_S]$ decays and (in view of expressions (3), (5) and $\lambda_{n\gamma} t_S \ll 1$) these decays give $\tilde{\nu}_e$ - flux:

$$S_1 = N_7^0 - N_7(t_S) - N_8(t_S) = \lambda_{n\gamma} N_7^0 t_S - (\lambda_{n\gamma} N_7^0 / \lambda_\beta) \varphi(V_S), \quad (6)$$

where we use the function

$$\varphi(y) = 1 - \exp(-\lambda_\beta y / w). \quad (7)$$

Nuclei of $^8$Li created within the previous time intervals decay in the same interval $[0, t_S]$ too. These previous intervals are: $[-V_0 / w, t_S - V_0 / w]$, $[-2V_0 / w, t_S - 2V_0 / w]$,..., $[-nV_0 / w, t_S - nV_0 / w]$, corresponding to the last, penultimate,..., "n"-th cycle with respect to the the instant $t = 0$. Taking into account that the nuclear concentration of the isotope $^7$Li does not practically vary, the antineutrino fluxes corresponding the considered decays are following:

$$S_2 = \frac{\lambda_{n\gamma} N_7^0}{\lambda_\beta} \varphi(V_B) \left\{ \exp\left[-\lambda_\beta(V_0 - V_B)/w\right] - \exp\left[-\lambda_\beta(V_0 - V_B + wt_S)/w\right] \right\}$$

$$S_3 = S_2 \exp(-\lambda_\beta V_0 / w) \quad (8)$$

.
.
.

$$S_n = S_2 \exp\left[-(n-2)\lambda_\beta V_0 / w\right].$$

Then the integral flux of lithium antineutrinos emitted from the volume $V_S$ during the time $t_S$ is:

$$N_S(t) = \frac{t}{t_S}\left(S_1 + \sum_{n=2}^{\infty} S_n\right) = \frac{t}{t_S}\left[S_1 + \frac{S_2}{\varphi(-\lambda_\beta V_0 / w)}\right]. \quad (9)$$

At a pumping over time $t_S = t_p$ the formula (10) gives an integral flux from whole volume of a blanket.

*2.2.2. Antineutrino flux from the delivery channel and pumped reservoir*

We shall neglect (n,γ)-activation of $^7$Li isotope in channels and in pumped reservoir. Let $t_d = L_1 / \mathcal{V}$ is the time of lithium delivery from the blanket to the pumped reservoir, i.e. the time which necessary for passage of the distance $L_1$ with linear velosity $\mathcal{V}$ (see figure 3).

As well as in case of a blanket, the integral flux of lithium antineutrinos from a channel of delivery during a time $t$ is a sum of infinite series of fluxes emitted from a channel within the time intervals $[t_p, t_p + t_d]$, $[t_p - V_0 / w, t_p + t_d - V_0 / w]$, ... , $[t_p - nV_0 / w, t_p + t_d - nV_0 / w]$:

$$N_{cd}(t) = \frac{t}{t_p} \sum_{n=1}^{\infty} \frac{\lambda_{n\gamma} N_7^0}{\lambda_\beta} \varphi(V_B) \varphi(wt_d) \exp\left[-(n-1)\lambda_\beta V_0 / w\right] = \frac{\lambda_{n\gamma} N_7^0 t}{\lambda_\beta t_p} \cdot \frac{\varphi(V_B)\varphi(wt_d)}{\varphi(V_0)}. \quad (10)$$

The expression (11) allows simply obtain the flux from the pumped reservoir. Let $V_R$ - volume of a pumped reservoir. The lithium antineutrinos flux from a reservoir is the remainder of fluxes $N_{cd}$ for two delivery times $[(t_d + V_R / w)$ and $t_d)]$:

$$N_R(t) = N_{cd}(t_d + V_R / w) - N_{cd}(t_d) = \frac{\lambda_{n\gamma} N_7^0 t}{\lambda_\beta t_p} \cdot \frac{\varphi(V_B)\varphi(V_R)\exp(-\lambda_\beta t_d)}{\varphi(V_0)}. \quad (11)$$

## 3. CHOISE OF PARAMETERS FOR LITHIUM ANTINEUTRINO SOURCE. RESULTS AND DISCUSSION

For the development of such source we need to realize the parametric analysis of characteristic for this installation. It is clear that the main requirements to the variable spectrum source are: high productivity factor $k$ of the lithium blanket; rather volumetric remote reservoir for maintenance of greater $\tilde{\nu}_e$-flux hardness in the location of a detector; rapid lithium delivery from the blanket to a reservoir, length of the lithium circulation loop must be minimized for the optimized $L_1$ (for example we have in mind the straight lengths $L_3$ must be short as possible); minimized length of the circulation loop allows to reduce the $^7$Li mass (figure 3). One of possible geometries is a system with a straight delivery channel. In this case the distance $L_1$ from blanket to reservoir will be maximal at the given delivery time $t_d$.

Let us define the relation between a whole system volume $V_0$, blanket volume $V_B$ and volume of the reservoir $V_R$ as follows:

$$\begin{cases} V_0 = (1+\alpha)V_B \\ V_R = \alpha \cdot b \cdot V_B \end{cases} \quad (12)$$

where $\alpha, b$ - some coefficients, $\alpha > 0$, $0 \leq b < 1$.

For choice of blanket operation parameters we need to consider the variation for relative number of lithium antineutrinos emitted from the blanket as function of a time of lithium pumping through the blanket $t_p$ and $\alpha$-coefficient. An information on operation parameters for a pumped reservoir and channels, the data for determination of relation between volumes of a blanket, reservoir and delivery channel can be obtained from dependence for relative number of $\tilde{\nu}_e$ emitted from the reservoir and delivery channel as function of: time $t_p$, coefficients $\alpha, b$, and delivery time $t_d$.

The assemblage of curves for relative number of lithium $\tilde{\nu}_e$ emitted from a blanket (i.e., $N_B(t)/N_0(t)$, where $N_0(t) = \lambda_{n\gamma} N_7^0 t$ - number of antineutrinos emitted from the whole system) as function of the coefficient $\alpha$ at various times of blanket pumping over $t_p$ (s) is given in figure 4.

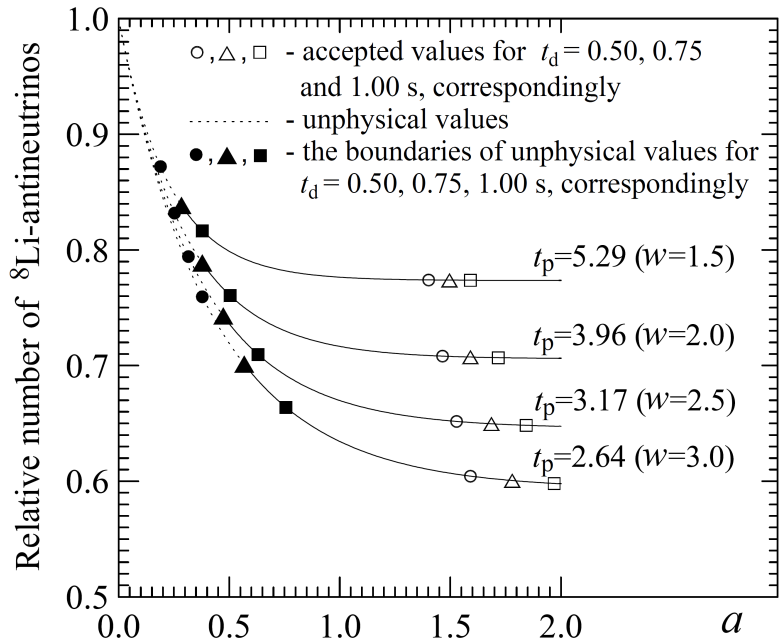

Figure 4. Relative number of lithium antineutrinos escaping from a blanket as function of the cyclic system volume parameter at different times of blanket pumping $t_p$ (s) [for flow rate $w$ (m$^3$s$^{-1}$)]. The accepted for calculation values correspond to the time of lithium delivery $t_d$ (s) from the blanket to the reservoir for the specified geometry.

The flow rates $w$ (m$^3$s$^{-1}$) (at which the given time of pumping over $t_p$ will be realized) are specified in brackets. The unphysical interval $\alpha = 0 \div \alpha_{min}$ can be found for the values $t_d$ and $w$ on the geometrical restrictions: $b = 0$ and $V_B(1 + \alpha_{min}) = (V_B + 2 w t_d)$. The boundaries of unphysical ranges are denoted at $\alpha_{min}$ values. The choice of possible values of delivery times $t_d$ is dictated by rapid $\beta^-$-decay of $^8$Li isotope. At large times of pumping $t_p$ the curves rapidly go on an asymptotic behaviour and further increase of $\alpha$ (volumes of reservoir and channels) with the purpose to increase a part of a hard lithium component in the total $\tilde{\nu}_e$- spectrum is unjustified. Therefore, the modes with a maximally possible flow rate $w$ are necessary.

The relative number of lithium antineutrinos emitted from a delivery channel $N_{cd}(t)/N_0(t)$ is a weak function of the coefficient $\alpha$ (see figure 5). The unphysical interval $\alpha = (0 \div \alpha_{min})$ is found under the same restrictions. The $\tilde{\nu}_e$ -part increases rapidly with growth of pumping rate and time of delivery $t_p$.

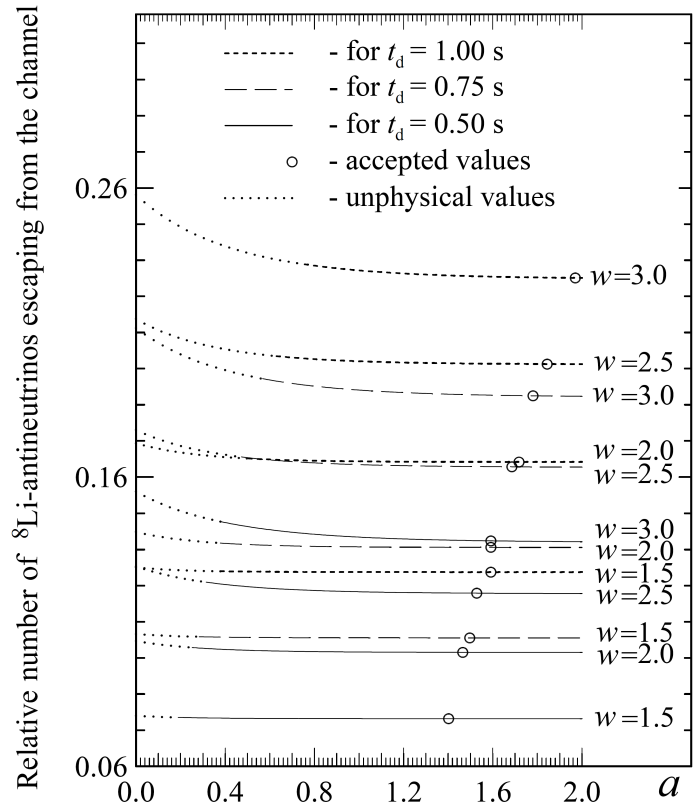

Figure 5. Relative number of lithium antineutrinos escaping from a delivery channel as function of the cyclic system volume parameter: at different values of the flow rate $w$ (m$^3$s$^{-1}$) and time of lithium delivery from the blanket to the reservoir - $t_d$ (s). The accepted for calculation values correspond to the time of lithium delivery $t_d$ (s) from the blanket to the reservoir for the specified geometry.

The assemblage of dependences for relative number of lithium antineutrinos escaped from a pumped reservoir [ i.e., $N_R(t)/N_0(t)$ ] as function of the coefficient $b$ at various times $t_p$ (the corresponding flow rates - see in the brackets), delivery times $t_d$ and parameters $\alpha$ are given in figure 6. The unphysical interval $b = (b_{max} \div 1)$ is found from the restriction: $\alpha V_B = (2 w t_d + \alpha b_{max} V_B)$. The specified $\alpha$-values in the figure 6 correspond to the considered geometries for which the calculations of the $\tilde{\nu}_e$ -spectrum hardness and neutrino reaction cross sections in the detector were made. It is seen that at larger times of pumping $t_p$ the part of lithium antineutrinos emitted from the reservoir rapidly goes on the asymptotic behaviour with growth of the reservoir volume. On the other hand, at growth of a pumping over rate $w$ the increase of a volume $V_R$ gives significant growth of the antineutrino part escaped from the reservoir.

For calculation of the generalized hardness and expected cross section in the total neutrino flux the next geometrical parameters were specified (figure 3): thickness of the spherical blanket $L_B = 1$ m; external diameter of the active zone - 48 cm (it corresponds to 51 liters volume of the high flux reactor PIK constructed now near the Saint-Petersburg); diameters of the channels - 0.40 m and it's turn - D= 2.70 m; length $L_3 = 0$. The considered blanket on the basis of heavy water LiOD solution with concentration $\geq 5.7$ % and 0.9999 purity of $^7$Li possess the productivity $k \geq 0.10$. The computations were made at different delivery channel lengths $L_1$ determined from the specified delivery time $t_d$ =0.5, 0.75, 1.0 s at fixed flow rate $w$ =1.5, 2.0, 2.5 and 3.0 m$^3$s$^{-1}$. The values corresponding just such geometries are marked in figure 4 - 6 as the accepted values.

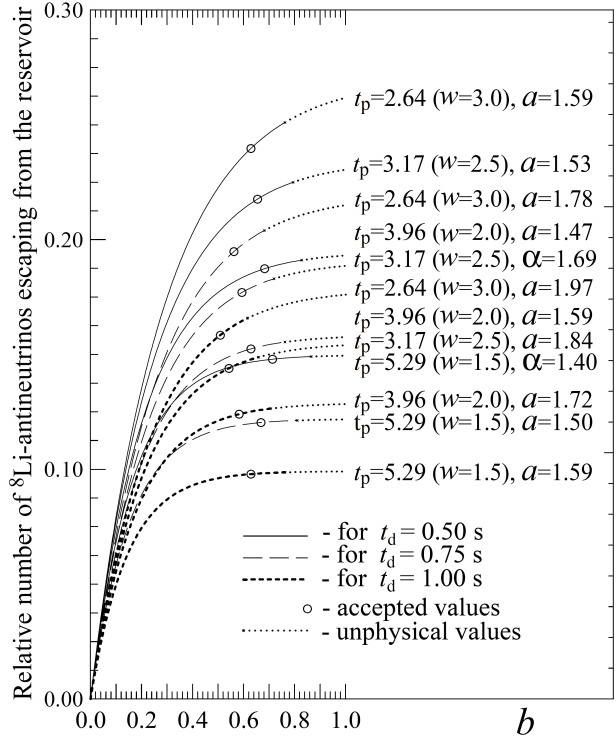

Figure 6. Relative number of lithium antineutrinos escaping from a pumped reservoir as function of the cyclic system volume parameter $b$ at different values of: time of converter pumping $t_p$ (s), cyclic system volume parameter $\alpha$, time of lithium delivery from the blanket to the reservoir $t_d$ (s). The accepted values correspond to the specified geometry.

For calculations of the total $\tilde{\nu}_e$ - spectrum hardness at the distance $L_2$ (averaged position of the remote detector) we need to integrate fluxes from small volume elements (according to the part 2.2 technique) of the whole system (figure 3) and taking into account the density of $^8$Li creation in the blanket. Dependence of the generalized hardness $H(\vec{r})$ as function of the distance $L_2$ at different flow rates $w$ were obtained for distances $L_2 = 2.5 - 30$ m. Results for hardness of the total $\tilde{\nu}_e$-spectrum at these distances $L_2$ for lithium delivery time $t_d = 0.75$ s are given in the figure 7. Values of flow rate $w$, corresponding to distances $L_1$ and velocity $\mathcal{V}$ are presented. These results are shown for the blanket productivity $k = 0.10$. For another coefficient $k'$ the value $H(\vec{r})$ should be multiplied on the ratio $(k'/k)$ as far as the productivity enters to the generalized hardness as a coefficient.

Cross sections of the $(\tilde{\nu}_e, d)$ reaction in the neutral and charged channels -

$$\tilde{\nu}_e + d \rightarrow n + p + \tilde{\nu}_e \quad \text{- neutral (n,p)-channel and} \tag{13}$$

$$\tilde{\nu}_e + d \rightarrow n + n + e^+ \quad \text{- charged (n,n)-channel,} \tag{14}$$

corresponding the reached hardness (figure 7) were calculated and presented on the the right axes taking into account the linear dependence of neutrino cross sections on the total spectrum hardness [4].

Actually, the total number of $\tilde{\nu}_e$ is:

$$N_{\tilde{\nu}_e} = N_{AZ} + H(\vec{r}) \frac{N_{AZ}}{\bar{n}}, \tag{15}$$

where $N_{AZ}$ - number of $\tilde{\nu}_e$ from the active zone, $\bar{n}_\nu$ - number of $\tilde{\nu}_e$ from the active zone per one fission. The second summand determines the number of lithium antineutrinos.

Then cross section $(\tilde{\nu}_e, d)$-reaction in (n,p)- and (n,n)-channels (per one fission) are also additive values:

$$\sigma_{np}(\vec{r}) = \sigma_{np}^{AZ} + H(\vec{r}) \times \sigma_{np}^{Li}, \tag{16}$$

$$\sigma_{nn}(\vec{r}) = \sigma_{nn}^{AZ} + H(\vec{r}) \times \sigma_{nn}^{Li}, \tag{17}$$

where cross sections from the active zone ($\sigma_{np}^{AZ}$, $\sigma_{nn}^{AZ}$) and from lithium blanket ($\sigma_{np}^{Li}$, $\sigma_{nn}^{Li}$) are calculated separately with their own spectrum. For calculation of the cross section in the total $\tilde{\nu}_e$-spectrum we used the data from [14,15].

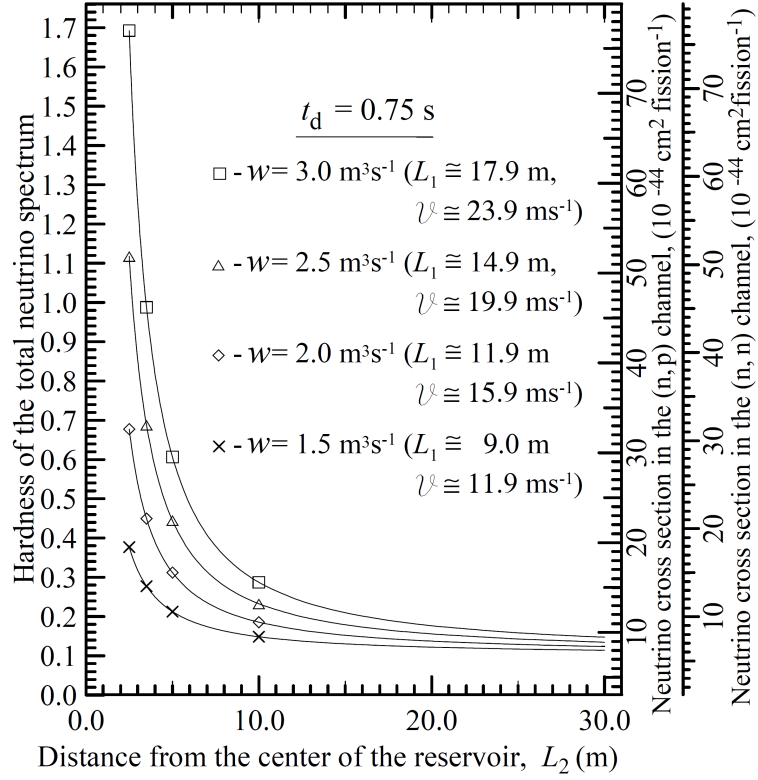

Figure 7. Generalized hardness $H(\vec{r})$ of the total $\tilde{\nu}_e$-spectrum and cross section of ($\tilde{\nu}_e$,d)-interaction with in (n,p)- and (n,n)-channels as function of the distance $L_2$ from the reservoir center. Curves are presented for different flow rate $w$ at fixed delivery time $t_d$ (s) from the blanket to the pumped reservoir.

Near by the reservoir the hardness rapidly drops with an increase of a distance $L_2$ and asymptotically tends to the value $H = 0.10$, - i.e. to the total $\tilde{\nu}_e$-spectrum hardness for the installation operating as a steady spectrum sources. For the fixed distance $L_1$ the greater hardness is reached at smaller distances from the reservoir $L_2$, and maximum flow rate $w$. In this way it is possible to increase cross sections in (n, p) and (n,n)-channels at the order and more in comparison with cross sections at $H = 0.10$ (when $w=0$, i.e., in a steady spectrum mode of source operation). In comparison with cross sections of these reactions in the purely reactor $\tilde{\nu}_e$-spectrum [7] the cross sections in the neutral channel of ($\tilde{\nu}_e$,d) interaction grow in tens times and in the charged channel - up to two orders.

In order to reach rapid pumping over of a blanket and to provide the lithium delivery on the distance $L_1 \approx 15 \div 25$ m in the time $t_d \leq 1$ s it is necessary to ensure a very significant flow rate $w$ and linear velocity $\mathcal{V}$ of moving in channels. An engineering of powerful pumps for such flows is possible and are realized at known reactors as Advanced Test Reactor (coolant flow rate up to 200 m³min⁻¹), Grenoble High Flux Reactor (velocity of heavy water coolant - 15.5 m s⁻¹), Savanna River High Flux Reactor (flow rate - 6.1 m³s⁻¹, linear velocity from 18.4 up to 21.5 m s⁻¹). In fact the increase of these values in the reactors are limited by use mixed and complicated geometries in the active zone (presence of many fuel rods which cause turbulence and so on). But in case of smooth surfaces of considered installation this problem is more simple and rate of flow can be ensured.

The proposed neutrino source with variable hard spectrum allows to modify a spectrum shape and investigate neutrino reactions at the different total $\tilde{\nu}_e$-spectrum hardness in the continuous interval from minimum to top value: $H = k \div H(w_{max})$. It can be achieved by varying flow rate $w$ from zero up to the maximum. This dependence of the generalized hardness and cross sections for the channels of ($\tilde{v}_e$,d) reaction are shown in figure 8 at the fixed length of the delivery channel $L_1 = 17.90$ m.

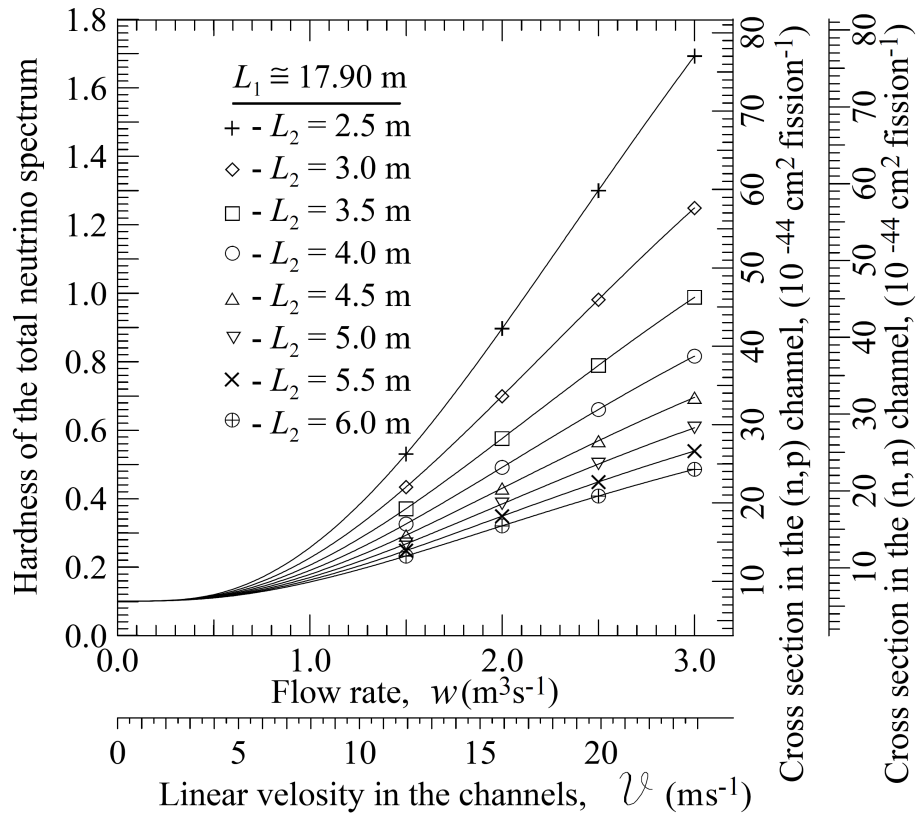

Figure 8. Generalized hardness $H(\vec{r})$ of the total $\tilde{v}_e$ - spectrum and cross section of ($\tilde{v}_e$,d) -interaction in (n,p)- and (n,n)-channels as function of the circulation rate $w$ for different distances $L_2$ from the reservoir center. The length $L_1$ of the delivery channel is fixed. Linear velocity $\mathcal{V}$, corresponding the flow rate $w$ is presented on the bottom axis.

The curves correspond to dependences for detectors position at the distances $L_2$ from 2.5 up to 6.0 m. The construction of such a facility requests 22.0 m$^3$ of lithium substance. In case of 5.7 % LiOD heavy water solution it mean ~0.5 t of $^7$Li isotope with purity 0.9999. The smooth growth of cross sections at increase of flow rates will give an opportunity to obtain more reliable experimental results.

The lithium antineutrino spectrum is attractive for oscillation experiments owing to well known distribution and it's hardness. These spectrum features can be especially helpful for search of sterile neutrinos. The problem of mass scale $\Delta m^2$ (between sterile and active neutrinos) is discussed intensively and some results indicate on eV-scale [16]. In this case the length of oscillation for $^8$Li antineutrinos will be about ~10 m and search of oscillation must be concentrated on short base line experiments [17,18]. Regimes with very fast lithium delivery and long distance $L_1$ can ensure large values of hardness $H$. In such experimental setup the advantages of the hard lithium spectrum ($\bar{E}_{\tilde{v}} = 6.5$ Mev) will answer the demands of short base.

## 4. REQUIREMENTS TO THE REACTOR AS A NEUTRON SOURCE FOR ACTIVATION OF THE LITHIUM BLANKET

It is necessary to discuss briefly the main requirements to the reactor used as neutron source for activation of $^7$Li isotope. This research reactor is used for production of neutrons escaping to the lithium blanket. The power generated in the active zone is a interfering factor of research reactors. But the purpose of creation of the high neutron fluence irradiating the outer blanket dictates the necessity of high level of power. The most often used fuel is enriched $^{235}$U. The other main fuel isotopes ($^{238}$U, $^{239}$Pu, $^{241}$Pu) are excluded and it simplifies the evaluation of the total neutrino spectrum at burning of the fuel. The one more advantage of the enriched $^{235}$U: the recently observed distortion of the reactor antineutrino spectrum in the 5-7 MeV can restricted by decay products only single fuel isotope [19].

The next demand to the discussed reactor is the compactness of it's active zone. In this case the the lithium blanket can be shifted more close to the reactor core. It means that the volume of $\bar{v}_e$-source will more compact too that is very important for precision of oscillation search. The such solution also allows to decrease significantly the requested mass of high purified $^7$Li. The possible examples for the discussed reactor are SM, HFBR and PIK research reactors (see table I).

Table I

Some parameters of high flux research reactors

| reactor | fuel | Volume of the reactor core, liters | Height of the core, cm | eeffective size of the core, cm | Max thermal power, MW | Max thermal neutron flux density | References |
|---|---|---|---|---|---|---|---|
| SM | $^{235}$U, enrichment - 90% | 50 | 35 | 42x42 | 100 MW | 5x10$^{15}$cm$^{-2}$s$^{-1}$ | [20] |
| HFBR | $^{235}$U (9.8 kg), higly enriched | ~25.4 | 53 | 48 (in diameter) | 60 MW | 1x10$^{15}$cm$^{-2}$s$^{-1}$ | [21] |
| PIK | $^{235}$U, enrichment - 90% | 50 | 50 | 39 | 100 MW | 5x10$^{15}$cm$^{-2}$s$^{-1}$ | [22, 23] |

**5. CONCLUSION**

An intense antineutrino source with variable (regulated and controlled) hard spectrum on the base of short lived $^8$Li isotope is discussed. The proposed source is characterized by not only hard spectrum but the unique feature - the possibility to vary hardness of the spectrum and to realize it smoothly without any changes in the installation.

The idea of the discussed source is to construct a closed loop where a high-purified $^7$Li isotope (or it's chemical compound) will be (n,γ)-activated (in the blanket) and pumped further to remote volume (close to the neutrino detector) and then will be pumped back to the blanket for the next activation. The operation of such source in a continuous cycle will ensure a hard antineutrino flux in the detector.

The hardness of the total $\tilde{v}_e$-spectrum can be tuned by regulation of lithium flow rate that ensure the possibility to vary the neutrino flux parameters (spectrum distribution and value of the flux) very quickly and without stop of the experiment.

The possibility to ensure very hard and variable antineutrino spectrum with well defined distribution can be very helpful in the short baseline experiment for search of sterile neutrinos with $\Delta m^2$ in the range about 1 eV$^2$ on which indicates the results of fits for some oscillation experiments

An analysis of the regimes are made parametrically (basing on the obtained analytical expressions for flux) that allows to evaluate the expected characteristics of the antineutrino source. The simulation demonstrated that owing to rise of generalized hardness the cross section of neutrino reaction increases significantly. So, the cross sections in (n,p)-channel of the reaction with a deuteron grow in tens times and in (n,n)-channel - up to two orders in comparison with these cross sections in the spectrum of reactor antineutrinos.

The continue of discussion for the item of hard $\tilde{v}_e$-source with variable and regulated (controlled) spectrum is given in the work [24].

**ACKNOWLEDGEMENTS**
The author thank Yu. S. Lutostansky for helpful and useful discussion.
The author tender thanks to L. B. Bezrukov, B. K. Lubsandorzhiev and I. I. Tkachev for their interest to this investigation and support of the work.

**REFERENCES**

[1] Huber P. *Determination of antineutrino spectra from nuclear reactors* // Phys. Rev. 2011. C **84** 024617.
[2] Aleksankin V. G., Rodichev S. V., Rubtsov P. M., Ruzansky P. A. and Chukreev F. E. *Beta and antineutrino radiation from radioactive nuclei* (Moscow: Energoatomizdat, 1989) ISBN 5-283-03727-4 (in Russian).
[3] Mikaelian L. A., Spivak P. E. and Tsinoev V. G. *A proposal for experiments in low-energy antineutrino physics* // Nucl. Phys. 1965. **70** 574-76.
[4] Lyutostansky Yu. S. and Lyashuk V. I. *Powerful hard-spectrum neutrino source based on lithium converter of reactor neutrons to antineutrino* // Nucl. Sci. Eng. 1994. **117** 77-87.
[5] Kopeikin V. I. *Flux and spectrum of reactor antineutrinos* // Phys. At. Nucl. 2012. 75 143-52.
[6] Lyashuk V. I. and Lutostansky Yu. S. *Intensive Neutrino Source on the Base of Lithium Converter.* arXiv:1503.01280v2 [physics.ins-det]. 2015.
[7] Lyutostansky Yu. S. and Lyashuk V. I. *Reactor neutrons-antineutrino converter on the basis of lithium conpounds and their solutions* // Sov. J. Atomic Energ. 1990. **69** 696–99.
[8] Lutostansky Yu. S. and Lyashuk V. I. *The concept of a powerful antineutrino source.* // Bull. Russ. Acad. Sci. Phys. 2011. **75** 468-73.
[9] Lyashuk V. I. and Lutostansky Yu. S. Neutron Sources foe Neutrino Factory on the Base of Lithium Convereter *Proc. of XXI International Seminar on Interaction of Neutrons with Nuclei (ISINN-21), Alushta, May 20-25, 2013* (Dubna: JINR, 2014) pp 156-64.
[10] Lyashuk V. I. and Lutostansky Yu. S. Neutron Sources for Neutrino Investigations (as Alternative for Nuclear Reactors) *Proc. of XXII International Seminar on Interaction of Neutrons with Nuclei (ISINN-22)* Dubna, May 27–30, 2014 (Dubna: JINR, 2015) pp 397-405.
[11] Lyashuk V. I. and Lutostansky Yu. S. *Powerful dynamical neutrino source with a hard spectrum* // Phys. Atom. Nucl. 2000. **63** 1288-91.
[12] Melnikova T. N. and Mozgovoi A. G. *Lithium isotope density and thermal-expansion* // High temperature. 1998. **26** 848-54.
[13] Lyashuk V. I. and Lyutostansky Yu. S. The Conception of the powerful dynamic neutrino source with modifiable hard spectrum. *Preprint ITEP-38-97.* Moscow: ITEP, 1997.
[14] Vidyakin G. S., Vyrodov V. N., Gurevich I. I., Kozlov Yu. V., Martem'yanov V. P., Sukhotin S. V., Tarasenkov V. G., Turbin E. V. and Khakimov S. Kh. *Observation of weak neutral current in interaction of fission antineutrinos with deuterons* // JETP Lett. 1989. **49** 279-81.
[15] Nakamura S., Sato T., Ando S., Park T.-S., Myhrer F., Gudkov V. and Kubodera K. *Neutrino-deuteron reactions at solar neutrino energies*. arXiv:nucl-th/0201062v3. 2002. http://www-nuclth.phys.sci.osaka-u.ac.jp/top/Netal/index.html
[16] Kopp J., Maltoni M. and Schwetz T. *Are There Sterile Neutrinos at the eV Scale?* // Phys. Rev. Lett. 2011. **107** 091801.
[17] Bungau A. et al. *Proposal for an electron antineutrino disappearance search using high-rate $^8$Li //* Phys. Rev. Lett. 2012. **109** 141802.
[18] Lyashuk V. I. and Lutostansky Yu. S. *Intense antineutrino source based on a lithium converter. Proposal for a promising experiment for studying oscillations* // JETP Letters. 2016. **103** 293–97.
[19] Hayes A. C., Friar J. L., Garvey G. T., Ibeling Duligur, Jungman Gerard, Kawano T., Mills Robert W. *Possible Origins and Implications of the Shoulder in Reactor Neutrino Spectra* arXiv:1506.00583v2 [nucl-th]. 2015.
[20] http://www.niiar.ru
http://www.niiar.ru/en/node/3004
[21] Shapiro S. M. *The high flux beam reactor at brookhaven national laboratory*, BNL-61645 http://www.iaea.org/inis/collection/NCLCollectionStore/_Public/26/074/26074008.pdf
[22] Konoplev K. A. PIK reactor state of construction 2007. *Proceedings of an International Conference organized by the International Atomic Energy Agency (IAEA), Sydney, 5-9 November 2007.* http://www-pub.iaea.org/MTCD/publications/PDF/P1360_ICRR_2007_CD/datasets/K.A.%20Konoplev.html

[23] Aksenov V. L. Reactor PIK. Present status and trends. *International Workshop "Collaboration and Perspectives of Russian and Chinese Mega Projects" December 3-4, 2014, Dubna, Russia.* http://lhe.jinr.ru/3-4DecWorkshop/files/Session%202/1_Aksenov.pdf

[24] Lyashuk V. I. *High flux lithium antineutrino source with variable hard spectrum. How to decrease the errors of the total spectrum ?* arXiv: 1612.08096 [physics.ins-det]. 2017.